\documentclass{aa}
\usepackage{graphicx}
\usepackage{txfonts}

\def\Vsigma{\ifmmode{V/\sigma}\else{$V/\sigma$}\fi}
\def\h3{\ifmmode{h_{3}}\else{$h_{3}$}\fi}
\def\Reff{\ifmmode{R_\mathrm{eff}}\else{$R_\mathrm{eff}$}\fi}

\begin{document}

\title{Shape and kinematics of elliptical galaxies: evolution due to
merging at $z < 1.5$}

\author{A. C\'esar Gonz\'alez-Garc\'{\i}a \and Jos\'e O\~norbe \and Rosa  Dom\'{\i}nguez-Tenreiro \and M. \'Angeles G\'omez-Flechoso}

\offprints{c.gonzalezgarcia@uam.es}

\institute{Departamento de F\'{\i}sica Te\'orica, Universidad Aut\'onoma de
  Madrid, 28049, SPAIN; c.gonzalezgarcia@uam.es}

\authorrunning{Gonz\'alez-Garc\'{\i}a et al.}
\titlerunning{Shape and kinematics of elliptical galaxies}

\date{Received <date> / Accepted <date>}

\abstract
{}
{
In this study we investigate the evolution of
shape and kinematics of
elliptical galaxies in a cosmological framework.
}
{We use a set of hydrodynamic, self-consistent
simulations operating in the context of a 
concordance cosmological model where relaxed
elliptical-like objects (ELOs) were identified
at redshifts $z=0$, $z=0.5$, $z=1$ and $z=1.5$.}
{The population of elliptical systems analysed
here present a systematic change through time,
i.e. evolution, by becoming rounder in general
at $z=0$ and, at the same time
more velocity dispersion supported. 
This is found to be primarily due to major
dry mergers where only a modest amount of angular momentum is involved
into the merger event.
Despite the general trend,
in a significant amount of cases the merger event
involves a higher specific angular momentum,
which in general causes the system to acquire a
higher rotational support and/or a more oblate shape.
These evolutionary patterns are
still present when we study our systems in
projection,  mimicking real observations, and thus they
should become apparent in future observations.}
{}


\maketitle

\section{Introduction}

Recent surveys of high-redshift galaxies suggest that some relaxed,
 massive elliptical galaxies could already be in place at $z \sim 2-1.5$ (Cimatti et al. 2004, Conselice et al. 2007), or even by $z \sim 5$ (Wiklind et al.~2008). 
 These results notwithstanding, according to the current formation scenarios, 
merging has played an important role in the mass assembly of most of the
local massive elliptical galaxies. In fact, observations and theory
suggest that, first, violent mergers at high $z$ have transformed most
 of the available gas into stars, and, later on,
these systems might have evolved through gas-free 
(the so-called dry) mergers (Conselice 2006;
Scarlata et al.~2007).

Since the study by Bertola \& Cappacioli (1975) we know that elliptical galaxies
are supported against gravity by random motions as well as rotation. Davies et al. (1983) studied the now classical $V_{max}/\sigma_o$ vs. $\epsilon$ diagram for spheroids (Illingworth 1977; Binney 1978), where $V_{max}$ is the maximum of the line-of-sight rotation curve, $\sigma_o$ is the central l.o.s. velocity dispersion and $\epsilon$ is the mean ellipticity inside a given radius.
They found that luminous (and massive) elliptical galaxies were characterised by low $V_{max}/\sigma_o$ and a fairly round aspect (low $\epsilon$), while ellipticals with intermediate luminosity tend to have larger values of $V_{max}/\sigma_o$ and $\epsilon$.
Several later observations in 1D spectroscopy of near-by elliptical galaxies 
(Lauer 1985, Bender 1988, Nieto et al. 1989, Bender et al. 1994, Pellegrini
2005, Lauer et al. 2005) and 2D spectroscopy show such behaviour for
ellipticals (see Emsellem et al. 2007, Cappellari et al. 2007, hereafter
CAP07). Recently van der Marel \& van Dokkum (2007) presented evidences of
evolution of the rotation support of elliptical systems since z=0.5. Present
formation schemes should explain these observations on the kinematics and
shape of elliptical galaxies and their possible evolution.

A number of recent N-body simulations of isolated galaxy mergers have dealt with the population of the classical diagram and the formation of boxy and disky objects 
(Naab \& Burkert 2003, Gonz\'alez-Garc\'{\i}a \& Balcells 2005, Gonz\'alez-Garc\'{\i}a \& van Albada 2005, Naab, Khochfar \& Burkert et al. 2006, Bournaud et al. 2005, Robertson et al. 2006, Cox et al. 2006, Gonz\'alez-Garc\'{\i}a et al. 2006, Jesseit et al. 2007, Naab \& Ostriker 2007). These studies indicate that mergers between disk galaxies tend to produce too large rotation when compared with present day massive elliptical galaxies. Besides, mergers between elliptical galaxies do reproduce the observed characteristics of massive ellipticals. 
Khochfar \& Burkert (2003), Kang et al. (2007) (and references there in)
present first attempts of semi-analytical modelling to address the issue
of kinematics and shape in early type galaxies. Naab et al. (2007) study
the formation of three massive galaxies from cosmological initial
conditions. However, a detailed and statistical analysis of the internal
kinematics and shape of objects formed in fully self-consistent
cosmological simulations and their possible evolution is, to the best of our knowledge, still missing and mandatory to provide a clear picture of the
mechanisms at play in the formation and evolution of present day E's.

In this paper we present results from self-consistent cosmological simulations where we have investigated the shape and kinematic evolution of elliptical-like objects (here after ELOs)
for several redshifts. For the present analysis we consider both three
dimensional (hereafter, 3D) and projected onto the sky data.
Other kinematic parameters recently proposed will be investigated in a fore-coming paper, but are out of the present study for the sake of simplicity.

\section{Methods and numerical experiments}

\subsection{Simulations}

We have run seven hydrodynamical simulations in the context of a
concordance cosmological model (Spergel et al. 2006). For five of the
simulations we employ a $10$ Mpc side 
periodic box with flat $\Lambda$CDM cosmological model, with $h=0.65$, $\Omega_{\rm m} = 0.35$ and $\Omega_{\rm b} = 0.06$.
To set the initial conditions we employed the algorithm developed by
Couchman (1991)  with an slightly high normalisation
parameter input $\sigma_{\rm 8}=1.18$,
as compared with the average fluctuations,
to mimic an active region of the Universe
(Evrard, Silk \& Szalay 1990). These simulations formed the A-sample in
O\~norbe et al. (2005, 2006, 2007) and they differ from each other in the
seed used to generate the initial conditions.

\begin{figure*}
\resizebox{\hsize}{!}
{\includegraphics[width=12cm]{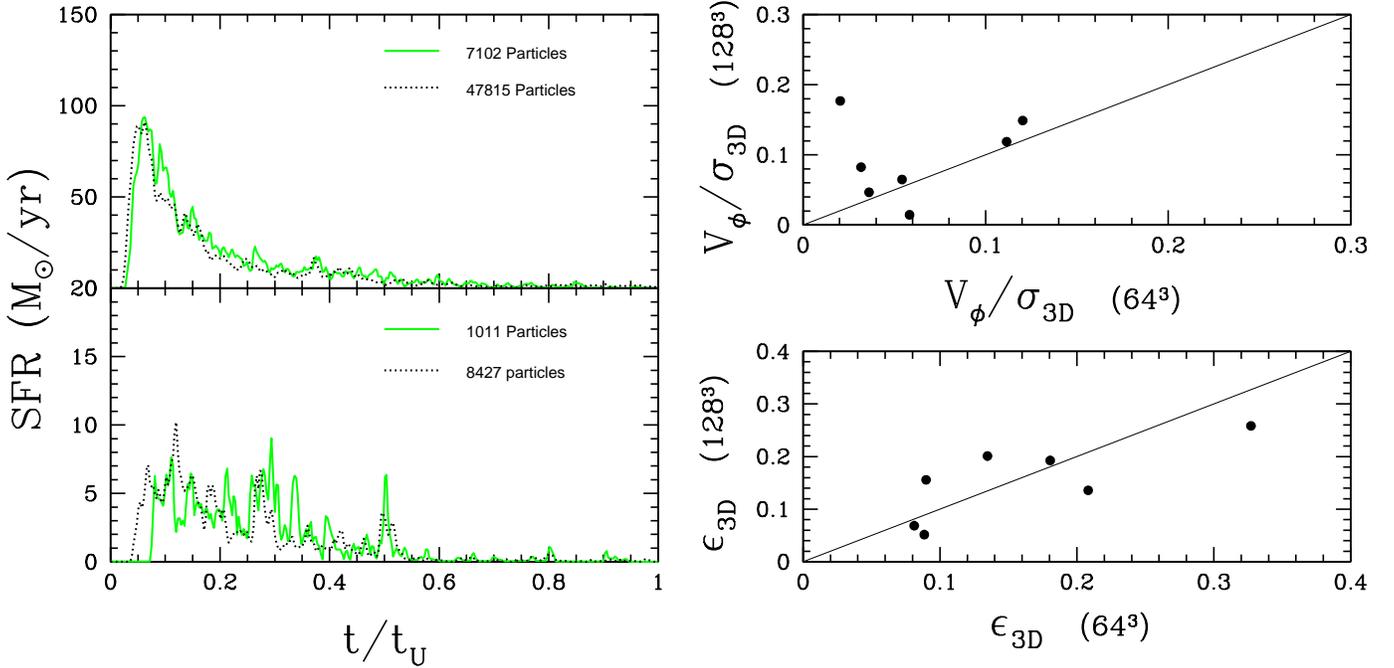}}
\caption{Left: Star formation history of the most (top) and least (bottom)
 massive ELOs for the test simulations with $2 \times 64^3$ particles
 (solid green line) and $2 \times 128^3$ particles (dotted black
 line). Right, comparison of the kinematic (top) and shape (bottom) observables. For details, see text. 
\label{fig:sfrt}}
\end{figure*}

In order to test if the evolutionary trends found in these 
simulations are robust, we have run two simulations
with a box side double as that in the first simulations and 8 times the number
of particles, with the same cosmology as before, and now using input values
$\sigma_8 =0.75$ and $0.95$. The first of these values is such that
the initial distribution of kinetic energy per unit volume is similar to that
in the small box simulations (Gelb \& Bertschinger 1994; Sirko 2005; Power \& Knebe 2006).
To run the small and larger box simulations,
we have respectively used DEVA 
(Serna, Dom\'{\i}nguez-Tenreiro \& S\'aiz, 2003)
and P-DEVA (its OpenMP version, see Serna et al., in preparation),
two Lagrangian SPH-AP3M codes. The mass resolution is $1.28 \times 10^{8} {\rm M_{\odot}}$ for dark matter particles and $2.67 \times 10^{7} {\rm M_{\odot}}$ for baryon particles.
Star formation (SF) processes have been included through a simple
phenomenological parametrisation (Katz 1992, see also
Serna et al. 2003 and O\~norbe et al. 2007 for details).

We have applied a consistency test by running two simulations with
identical initial conditions in a $10 \rm{Mpc}$ side periodic box, but one 
has $2 \times 64^3$ particles while the second has $2 \times 128^3$
particles, with masses as above for the  $2 \times 64^3$ simulation and
$1.6 \times 10^{7} {\rm M_{\odot}}$ for dark matter particles and $3.3
\times 10^{6} {\rm M_{\odot}}$ for baryon particles for the $2 \times
128^3$ simulation.

Galaxy-like objects of different morphologies appear in the simulations.
 ELOs have been identified
 as those having a prominent, relaxed spheroidal component,
made out of stars, with no extended discs. These baryonic objects are embedded in a dark matter extended halo, typically 10 times more massive. In O\~norbe et al. (2005, 2006)
it is shown that these ELOs satisfy dynamical Fundamental Plane relations. Also
in O\~norbe et al. (2007) the total, bright and dark matter profiles of ELOs
and their kinematics are analysed showing a very satisfactory agreement
with observational data.

\subsection{Methods and analyses}

We have analysed the ELOs at four redshifts: $z=1.5$, $z=1.0$,  $z=0.5$
 and $z=0$. We have selected undisturbed ELOs (i.e. within their
 limiting radius we observe a quasi-equilibrium behaviour of the system) because we want to analyse shapes, although their dark matter halos could be at the first stages of merging. We have analysed those systems that at each redshift have more total stellar mass than $2.6~{\times}~10^{10} {\rm M}_{\odot}$.
 Such limit is in agreement with Kauffmann et al. (2003) for early type
 galaxies, thus the number of particles per ELO is large enough to avoid
 resolution problems (see below). In total we have 425 ELOs
 (225 from the small box simulations and 200 from the big size
 simulations), 127 at $z = 0$, 107 at $z = 0.5$, 97 at $z=1.0$ and 94 at
 $z = 1.5$. 

To quantify the shape of the ELO we measured the axis ratios ($c/a$ vs. $b/a$, with $a > b> c$) of the ellipsoidal figures defined by the particles inside $r_{90}$ (the radius
enclosing $90\%$ of the stellar mass of the system), by computing the eigenvalues of their inertia tensor (see Gonz\'alez-Garc\'ia \& van Albada 2005). 
To classify the shape of our objects we use the triaxiality parameter introduced by de Zeeuw \& Franx (1991): $T=(1-(b/a)^2)/(1-(c/a)^2)$ in the following way: we introduce the parameter $s$ as $s=3$ if $c/a > 0.9$ (close to spherical objects), $s=2$ if $c/a < 0.9$ and $T < 0.3$ (oblate objects), $s=0$ if $c/a < 0.9$ and $T > 0.7$ (prolate objects) and $s=1$ elsewhere (triaxial objects). 
Concerning kinematics, we have studied the three dimensional velocity dispersion inside  the effective radius $r_{e}$\footnote{$r_{e}$ is the radius enclosing half of the stellar mass of the system} ($\sigma_{3{\rm D}}$), and we have looked at the mean tangential velocity at $r = r_{e}$, ($V_{\phi}$).
We have also looked at these characteristics at $r_{90}$.

To compare with classical observations, we have investigated the  line-of-sight rotation and velocity dispersion curves up to one projected effective radius ($R_e$) for each object. 
We chose a point of view perpendicular to the spin angular momentum vector of
the stellar matter because these particular points of view should maximise the effects of rotation when these are present (see Binney 2005, Burkert \& Naab 2005).
We first derive the ellipticity of each ELO in the sample by
projecting the whole particle distribution onto a plane
perpendicular to the line-of-sight. Then, local surface densities were
computed from a logarithmic binning of space. Ellipses are fitted to the so
obtained isophotes, as it is customary done in observations. The ellipticity
computed in our analysis is the mean of the ellipticities inside one
$R_e$. Although the number of particles could be not very high in some
objects, the signal at half the mass radii in projection is high enough to
calculate these values. We have performed consistency tests with
methods based on a diagonalization of the moment-of-inertia tensors, and values are consistent to a good
level of agreement. To derive the line-of-sight rotation and velocity
dispersion curves, we have placed a slit along the major axis of the projected
system (as obtained in the previous step) and we have projected the velocities
of each particle along the line of sight. From these curves, we finally got 
the central line-of-sight velocity dispersion ($\sigma_o$) and the maximum of the velocity curve inside $R_e$ ($V_{max}$).

To understand the physical processes underlying the shape and rotation support changes, for each of the 127 ELOs at $z = 0$ we have drawn its mass aggregation track (MAT) along the main branch of its merger tree, both for baryonic (the mass inside fixed radii) and for its total mass (the virial mass).
These MATs give us information on the mass assembly processes through
time. Major mergers (MM; $M_{secondary}/M_{primary} > 0.25$), minor
mergers (mM) and aggregation (i.e. smooth in-fall of mostly gaseous
material) processes can be clearly identified. We can also compute the amount of dissipation involved in the different processes as well as estimate the amount of angular momentum involved.
 Complementary informations about a merger is provided by 
the configurations shown, in a time interval around the merger event, 
by the baryonic particles destined to form the ELO later on:
type and number of objects involved, environment.
This way we have analysed 150 events of any type 
along the main branch of the merger tree at $z < 1.5$, and classified them 
into several categories: 
mM or MM; binary or multiple;
involving a high or a low amount of
specific angular momentum (and their combinations).

\subsection{Consistency checks}

Figure~\ref{fig:sfrt} shows the results for the consistency
simulations. Fig.~\ref{fig:sfrt} left panel shows the star formation
history for two objects, the most and least massive ELOs in these
simulations. Black dotted line and solid green line depict the results
from the simulations with more and less particles respectively. We find
small differences, specially at early times, however the two systems
display general similar behaviour, and at high cosmic times (low
redshift) no big differences are noticed. A similar test was done by
Naab et al. (2007) and, although the numerical approaches are different, 
it is reassuring to find also a convergence in this resolution test. 

Figure~\ref{fig:sfrt} right presents comparative results for the same
objects at $z=0$ in the two simulations where we have computed the observables
introduced above on shape and kinematics. The systems behave quite well
and the agreement is good to a nice level, although there is one object
presenting a significant difference. This system is not the least massive in
these simulations and the difference is due to the peculiar way to measure
the $V_{\phi}$ parameter, that may include slightly different
particles. The overall consistency of these tests prompt us to
use the $2 \times 64^3$ simulations for the sake of computing time.

\section{Results}

\subsection{Shape and kinematic evolution}

\begin{figure*}
\resizebox{\hsize}{!}
{\includegraphics[width=120cm]{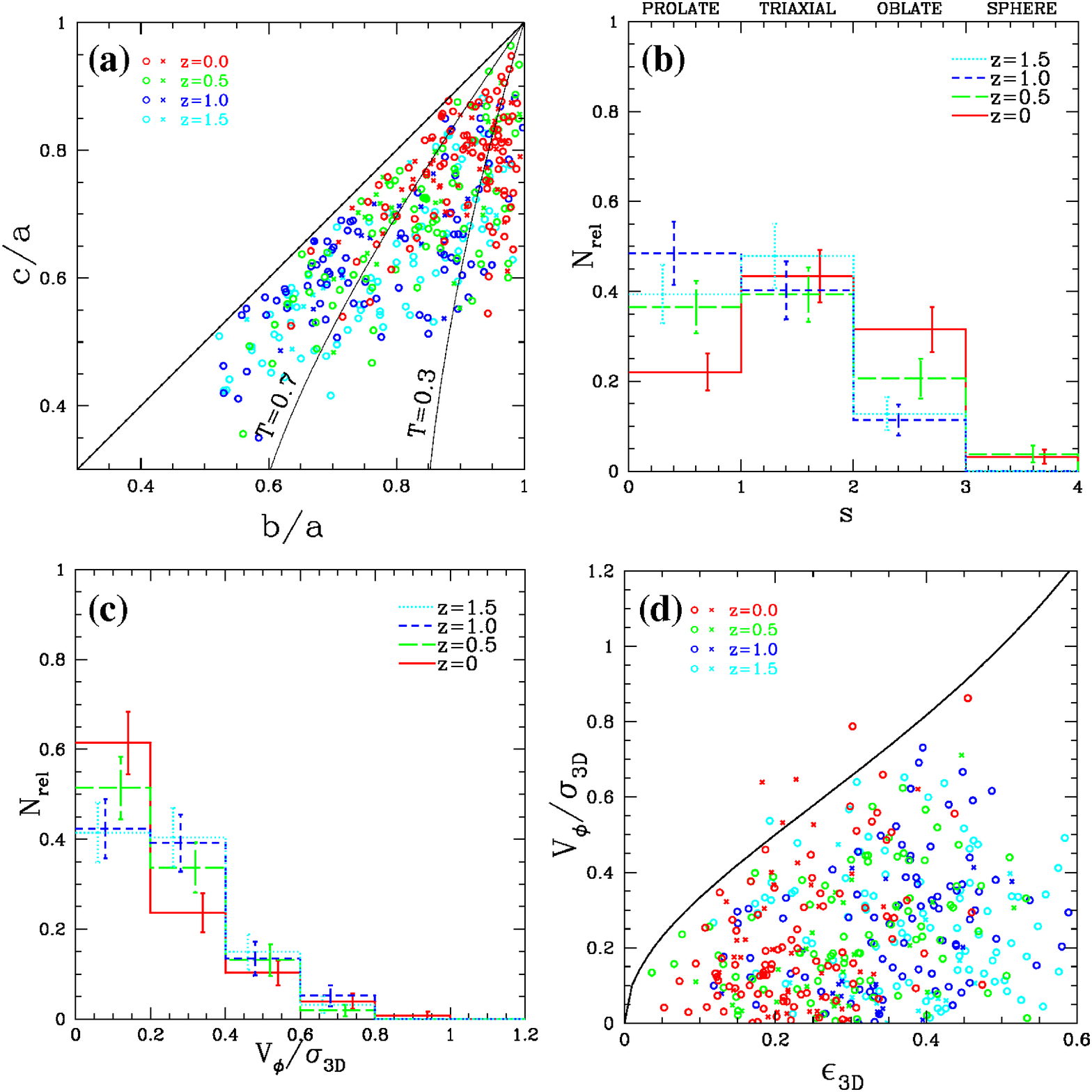}}
\caption{(a) Axis ratios of the spheroids. Colours indicate the different redshits. Crosses are objects more massive than $M_* > 1 {\times} 10^{11} {\rm M}_{\odot}$, circles are objects with less mass. (b) Histograms of the shape parameter $S$ vs. relative number for all our systems. Different lines and colours give the histogram for the four redshift bins. (c) Histograms of the rotational support measured as $V_{\phi}/\sigma_{3{\rm D}}$ (at $r_e$) in relative number for our systems, colours and lines as in (b). (d) $V_{\phi}/\sigma_{3{\rm D}}$ vs. $\epsilon_{\rm 3D}$ for the ELOs, colours and symbols as in (a) panel. Black solid line indicates the locus for oblate rotators. For details, see text. 
\label{fig:1}}
\end{figure*}

Figure~\ref{fig:1}, top row, shows the shape for the 4 redshifts,  presented in different colours. Fig.~\ref{fig:1} (a) shows $c/a$ vs. $b/a$. 
Blue symbols ($z > 1$) appear mostly in the middle-left of the diagram, while red symbols ($z = 0$) in the top right, 
that is, the values of $c/a$ and $b/a$ tend to increase with decreasing redshift, or in summary, the population of ELOs become rounder on average with time. These results hold when we use  $r_e$ instead of 
 $r_{90}$ to calculate  inertia tensors,
 although in this case ELOs tend to appear even rounder.
Fig.~\ref{fig:1} (b) shows a histogram for the shape parameter $s$ defined
above. Results are given in relative numbers, and the error bars express the Poisson noise. Most of our systems are triaxial objects at all redshifts, except at $z > 1$ when there is a fair amount of prolate systems. 
For all the simulations we have considered here,
the fraction of prolate objects decreases with decreasing redshift, and at all redshifts there is a very small number of perfect spheres.  Moreover, the number of triaxial and oblate objects increases with decreasing redshift. It is noteworthy that our results on ELO shapes at $z = 0$ compare reasonably well with those obtained  from the SLOAN survey for elliptical galaxies (Kimm \& Yi 2007),  where they show that there is about a $45\%$ of triaxial objects and around $26-29 \%$ of prolate and oblate objects in that sample. 

Fig.~\ref{fig:1} (c),
shows the rotational support for ELOs. The histograms are normalised to their total number at each redshift, and the error bars show the Poisson statistics. 
There is a trend towards increasing the number of 
systems with a lower value of $V_{\phi}/\sigma_{3\rm{D}}$ as $z$ decreases;
this trend appears independently from the details of the simulation.
Fig.~\ref{fig:1} (d) shows the diagram of $V_{\phi}/\sigma_{3\rm{D}}$ vs. $\epsilon_{\rm 3D}$, where $\epsilon_{\rm 3D}= 1-c/a$. Again, blue objects ($z > 1$) appear mostly at high $\epsilon_{\rm 3D}$ values, with a broad range of $V_{\phi}/\sigma_{3\rm{D}}$ values, while at $z = 0$ (red symbols) $\epsilon_{\rm 3D}$ and $V_{\phi}/\sigma_{3\rm{D}}$ are smaller on average.
These trends still hold  if we use $r_{90}$ instead of $r_{e}$
(i.e., we measure external rotation instead of rotational support)
although in this case there is a higher fraction of systems with larger values of $V_{\phi}/\sigma_{3\rm{D}}$.

 We have performed Kolmogoroff-Smirnoff tests to check whether the samples at
different $z$'s come for the same distribution. Such null hypotheses
can be ruled out  in all cases, both for shapes and support against gravity,
with confidence levels between 95 and 99.9 $\%$. Given the consistency on the direction of the more frequent changes (i.e., towards becoming 'rounder' and more pressure supported) we dare to call these changes an evolutionary track. We must emphasise that this evolution concerns the global population of E's rather than individual ones.

We now turn to 2D analysis. First of all, there is a reasonable agreement between the 3D and the projected kinematics, i.e. the low and fast rotating systems are in agreement in both samples. Also, objects with large $c/a$ ($\sim 1$) tend to be those with the smaller $\epsilon$, while the larger ellipticities are found among those objects with lower $c/a$. This correlation between 3D and 2D results for ELOs is noteworthy because it implies that the intrinsic evolution detected in virtual systems, must also be detected in observational (projected) data, should it occur in real ellipticals.

\begin{figure}
\resizebox{\hsize}{!}
{\includegraphics[width=8cm]{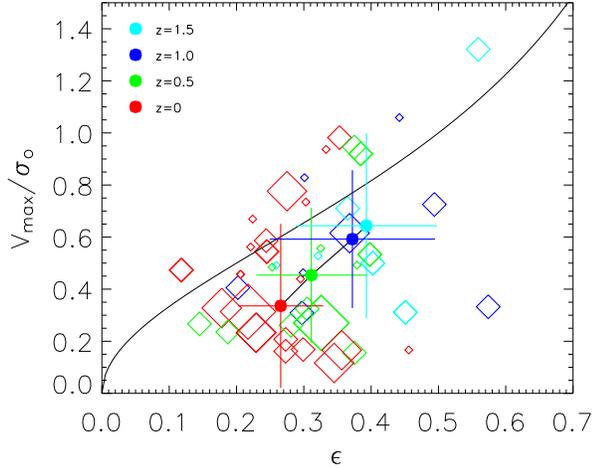}}
\caption{$V_{max}/\sigma_o$ vs. $\epsilon$ diagram. The black solid curve gives the locus of the oblate rotators (Binney 1978). Only models with $M_* > 1 {\times} 10^{11} {\rm M}_{\odot}$ are plotted. Colours are as in Fig~\ref{fig:1}. Filled circles give a mean for the objects at each redshift. Finally the size of the symbol gives the accumulated number of major mergers that a system has undergone.
\label{fig:3}}
\end{figure}

Figure \ref{fig:3} presents the results for  the classical $V_{max}/\sigma_o$
vs. $\epsilon$ diagram (Davies et al. 1983). For the sake of clarity, we plot
only the most massive objects $M_* > 1 {\times} 10^{11} {\rm M}_{\odot}$ of
the small-sized simulations and the values are measured from a point of view
perpendicular to the angular momentum vector. Each object is represented by an
open symbol, the colour represents the redshift for that object. At high {\it z} (light and dark blue), ELOs tend to appear as more flattened systems with a large relative rotation. However at lower redshifts (green and red) we have more massive systems. These tend to appear rounder and more pressure supported (except those few with large $V_{\phi}/\sigma_{3\rm{D}}$), in agreement with the 3D data (see Fig.~\ref{fig:1}(d)). The filled circles give the mean of the distribution for each redshift and the error bars are a measure of the standard deviation. The systems evolve on average from flatter and more rotationally supported towards rounder/triaxial and pressure supported systems. 
Finally, it is worth noting that if we include all systems in this plot, the evolution still holds. The size of the open symbols in Fig.~\ref{fig:3} is a measure of the accumulated number of major mergers that the system has suffered since $z=1.5$. Larger symbols are mostly located in the lower left part of the diagram. This indicates that merging is one of the key ingredients driving this evolution 
 towards rounder and more pressure supported ELOs.
Although this is the general trend, we also note that there are some cases with large
$V_{max}/\sigma_o$ and a fair amount of mergers, and this indicates that
merging effects could be far more complex. 

It has been recently argued (Binney 2005, Burkert \& Naab 2005, CAP07) that $V_{max}/\sigma_o$ vs. $\epsilon$ diagrams are affected by projection effects. The results presented in Fig.~\ref{fig:3} are obtained by looking at each object from a point of view perpendicular to the angular momentum vector. For oblate rotators such vector should be close to the small principal axis, and the projected image would provide the maximum ellipticity and $V_{max}/\sigma_o$.
However, many of our most massive systems are triaxial and the angular
momentum vector does not need to coincide or be near the short principal
axis. In that cases projection effects in our diagram could still be
present. To test the effects of projection we have looked at the systems
from 100 randomly chosen points of view and we have obtained the
projected observables of the classical diagram. We obtain a distribution
of points, which in the case of the most massive ELOs does not differ
greatly from the points presented in Fig.~\ref{fig:3}. We have tested
the null hypothesis that the distributions of points for the different
redshifts come from the same distribution through a Kolmogorov-Smirnoff
test. The result is that we can rule out such hypothesis to the $99.9\%$
confidence level. The same result is obtained through a Kuiper test. We
then conclude that the results presented are robust and that the
evolution observed in the projected quantities in shape and kinematics
is indeed real.

\subsection{Merging rate}

To understand how merging affects evolution,
 we have carefully studied  and classified all the merger events along 
the MATs of the 127 ELOs identified at $z=0$.

\begin{figure}
\resizebox{\hsize}{!}
{\includegraphics[width=8cm]{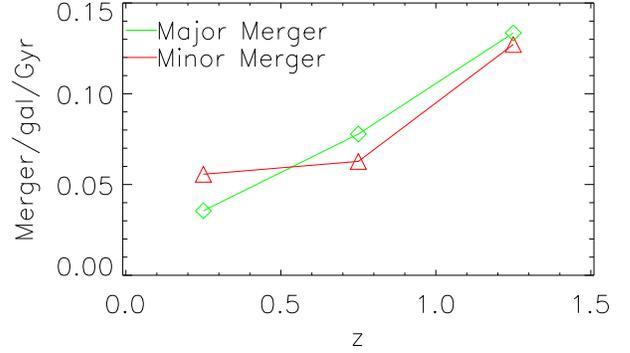}}
\caption{Number of mergers per galaxy per Gyr as a function of redshift
 interval. Red line gives the rate for minor mergers. Green line gives
 the rate for major mergers. We observe a decline in the number of
 mergers, this decline is slightly steeper in the fraction of MM.
\label{fig:4}}
\end{figure}

We have investigated the number of mergers in three redshift
intervals $z \in (1.5,1)$; $(1,0.5)$ and $(0.5,0)$. We are interested in
the mergers suffered by our ELOs. We 
investigate the main branch of the ELO merger tree and look at the mass
aggregation track to identify increments of mass. Thus, we identify
major mergers (MM) 
and minor mergers (mM). For the three redshift intervals we have identified
21, 23 and 23 MMs, and 20, 22 and 36 mMs. Figure~\ref{fig:4} shows the
evolution of the merger rate since $z = 1.5$ for the three redshift bins. The merger rate is
expressed in terms of the number of mergers per galaxy per gigayear. We
find a decrease in the number of mergers. This decrease is more
pronounced in the MM fraction than in mM. 

We must note that this result involve the number of
mergers suffered by the ELOs in our sample only along the main branch of
their merger tree. Besides, we analyse the high mass end of the galaxy
mass distribution, not taking into account
mergers resulting in disk-like 
systems or systems with smaller mass. Despite these caveats, we can
compare these numbers with recent theoretical results. Naab et al. (2007) perform
cosmological simulations including gas, cooling and star formation to
investigate the formation of three systems, two resemble true
elliptical galaxies while the third resembles and S0 galaxy. They find
that for $z < 1$ the systems have underwent few merger events, with one
major merger in one case and a minor merger in another one. Scannapieco
et al. (2008) use results from the Aquarius simulations (Springel et
al.~2008) to investigate the survival of disks. One important aspect is
the number of mergers in order to estimate their impact in the survival
of the disks. They find that the disks have suffered a moderate number
of mergers since $z=2$. Finally Genel et al. (2008, 2009) using the Millennium
simulation find that since $z=2.2$ the number of MM for a halo with mass
$M$ in the 
redshift interval $z_i$ and $z_f$ can be approximated by
$\bar{N}_{MM}(z_i,z_f,M) \approx 0.13\{\log[M/(10^{10} M_\odot)]+1\}(z_i
- z_f)$. Our simulations display slightly larger values of this
parameter, perhaps related to the active conditions modelled in our
runs. However, we find that the results are in agreement with Figure 8
in Genel et al. (2009) and our numbers are consistent with these
theoretical works. Finally, Conselice et al. (2009) study the merger
history of a large number of galaxies for $z < 1.2$. Our results here
are again in good agreement with those presented in their Fig.7 for
observational results.

\subsection{Trends of merger characteristics with time}

At high $z$ ($z > 1.5$)
 dissipative, multiple MMs are  more frequent
than other merger types. 
 At $z < 1.5$ most mergers are rather dry  
(ELOs are mostly devoid of gas, with percentages of less than a few percent), 
with
 a higher frequency of low angular momentum MM compared to 
high angular momentum MMs (a frequency of $ \sim 2:1$).
 Most mergers involving a fair amount
of specific angular momentum are multiple events.
However, the opposite is not true, as there are  multiple events 
where only a small amount of angular momentum is involved,
and particularly so at $z > 1.5$.

\section{Discussion}

We have  compared the merger characteristics with the changes measured
in shape and rotational support. We have found that there is a close correlation
between the shape and the amount of rotational support of a given ELO at a given time and the characteristics of the last merger event it has suffered. 
Specifically: MMs always produce changes in both shape and rotational
support in the outer and the inner parts.
MMs with little angular momentum will most often decrease the rotational support inside $r_{e}$ producing mostly rounder (i.e., a larger value of $c/a$) prolate spheroids, while
 MMs with a high (intermediate) amount of angular momentum 
increase rotational
 support producing
 oblate or triaxial systems. 
This picture  about the role of angular
momentum is consistent with the transformation found in binary  mergers of
spheroidal systems by Gonz\'alez-Garc\'{\i}a \& van Albada (2005).
mMs most often produce triaxial or oblate spheroids and 
increase the rotational support, although the effect is most likely to affect 
the outer parts of the ELO (see Balcells
\& Quinn 1990, Eliche-Moral et al. 2006), except for penetrating mMs.
Finally, aggregation processes are also important and may affect the shape and kinematics of the final object. 
We find that around $85\%$
 of the changes in shape and in rotational support are associated to merging
(as opposite to aggregation).

Combining these findings on the trends of merger characteristics with time
and the different effects a merger causes according to these characteristics, 
we can understand that prolate ELOs are the most frequent at high $z$,
and then they are transformed most often into rounder triaxial objects
with less rotation support. 
 An interesting conclusion from this study is
that, due to dry merging, the most luminous (i.e. massive) ellipticals must be rounder and less
rotationally supported in general. Indeed, this is what is
found in observational data (see e.g. Davies et al. 1983).

The results presented  here do not intend to fully represent 
at a quantitative level the  characteristics 
of the shape and kinematic evolution of elliptical galaxies, 
but rather to unveil some qualitative trends
in dense environments and the mechanisms
causing them in a cosmological context.
It is reassuring to find that all the simulations
analysed here show in general an evolution of the population of E's towards roundish and less rotationally supported systems, mostly driven by dry merging
 and that the mechanisms at play described above, causing either
the general trend or the exceptions, are consistent in all kind of simulations, run with different box size, particle number, input $\sigma_8$ and code.

Van der Marel \& van Dokkum (2007) report a similar evolution for two samples of $\sim 40$ galaxies at $z=0.5$ and $z=0$. Although the evolution we observe in our simulations seems milder than the one reported by this study, it remains open the amount of evolution to larger redshifts.
Thus, it would be highly desirable to perform larger statistical
studies of the shape and kinematics of real E's at low $z$ and to aim at 
observations
of kinematics at higher redshifts to confirm the evolutionary trend of ellipticals.

This work was partially supported by the MCyT (Spain) grants:
 AYA-07468-C03-03 and AYA2006-15492-C03-01 from the PNAyA, and the
 regional government of Madrid through the ASTROCAM Astrophysics network
 (S­0505/ESP­0237). We thank the Centro de Computaci\'on Cient\'ifica
 (UAM, Spain) for computing facilities.

\end{document}